\documentclass{ws-ijmpd}
\usepackage[super,compress]{cite}
\usepackage{graphicx}
\usepackage{epstopdf}
\usepackage{amsfonts}
\usepackage{amssymb}
\usepackage{epsfig}
\usepackage{hyperref}

\begin{document}


%
\catchline{}{}{}{}{}
%

\title{Quasinormal modes of black holes in Einstein-power-Maxwell theory}

\author{Grigoris Panotopoulos}

\address{Centro Multidisciplinar de Astrof\'{\i}sica, Instituto Superior T\'ecnico,
Universidade de Lisboa, Av. Rovisco Pais, 1049-001 Lisboa, Portugal
\\
\href{mailto:grigorios.panotopoulos@tecnico.ulisboa.pt}{\nolinkurl{grigorios.panotopoulos@tecnico.ulisboa.pt}} 
}

\author{{\'A}ngel Rinc{\'o}n}

\address{Instituto de F{\'i}sica, Pontificia Universidad Cat{\'o}lica de Chile,
Av. Vicu{\~n}a Mackenna 4860, Santiago, Chile
\\
\href{mailto:arrincon@uc.cl}{\nolinkurl{arrincon@uc.cl}} }

\maketitle

\begin{history}
\received{Day Month Year}
\revised{Day Month Year}
\end{history}

\begin{abstract}
In the present work we compute the spectrum of quasinormal frequencies of four-dimensional charged black holes in Einstein-power-Maxwell theory.
In particular we study scalar perturbations adopting the 6th order WKB approximation. We analyze in detail the behaviour of the spectrum depending on the charge of the black hole, the quantum number of angular momentum and the overtone number. In addition, a comparison is made between the results obtained here and the results valid for charged black holes in other theories as well as for the Reissner-Nordstr{\"o}m black hole. Finally, we have provided an analytical expression for the quasinormal spectrum in the eikonal limit.
\end{abstract}


\ccode{}

\section{Introduction}

General Relativity (GR) and Quantum Field Theory comprise the two cornerstones of modern theoretical physics. Within the framework of GR black holes (BH) as well as gravitational waves (GW) are predicted to exist. Stability of the exterior spacetime of classical BHs against small perturbations and quasinormal modes (QNM) have been an old issue in the field \cite{wheeler,zerilli1,zerilli2,zerilli3,moncrief,teukolsky} (see also the Chandrasekhar's monograph \cite{monograph}). BH perturbations generically are characterized by three stages. First, production of radiation due to the initial conditions, then damped oscillations (with complex frequencies), and finally a power-law decay of the fields. The complex frequencies of the form $\omega=\omega_R-\omega_I i$ depend entirely on the few BH parameters, namely the mass, the electric charge and the rotation speed, and are the so-called QNM of the BH. The real part determines the period of the oscillation, $T=2 \pi/\omega_R$, while the imaginary part describes the decay of the fluctuation at a time scale $t_D=1/\omega_I$.

Until recently there were only weak evidences regarding the existence of BH and GW, such as supermassive BH in the center of galaxies, and decay orbit
of binary systems due to gravitational radiation. Things have changed after the historical LIGO direct detection of gravitational waves \cite{ligo1,ligo2,ligo3} in BH mergers, which provides us with the strongest evidence so far that BHs exist and merge, and opens a completely new window to our Universe. The black hole perturbation, such as in the late stage of a black hole formation, can be described by quasinormal modes, and the gravitational radiation is expected
as a damped sinusoidal waveform. Since the spectrum of quasinormal frequencies carries unique information about the parameters of the black hole, by observing the quasinormal spectrum, i.e., the frequencies and damping rates, we can determine the mass and angular momentum of a spinning black hole. It thus becomes clear that studying QNMs is important, and it is for this reason that QNM of BHs have attracted a lot of attention lately. For an old review on QNM see e.g. \cite{review1}, while for a recent review on the subject see \cite{review2}.

Apart from the standard Maxwell's linear electrodynamics, other theories of electrodynamics, non-linear in nature, have been introduced and investigated in the literature for several reasons. Originally the Born-Infeld non-linear electrodynamics was introduced in the 30's in order to obtain a finite self-energy
of point-like charges \cite{BI}. Over the last decades this
type of action reappears in the open sector of superstring theories \cite{ST1,ST2} as it describes the dynamics of D-branes \cite{Dbranes1,Dbranes2}.
On the other hand, straightforward generalization of Maxwell's theory leads to the so called Einstein-power-Maxwell (EpM) theory described by a Lagrangian density of the form $\mathcal{L}(F) = F^k$, where $F$ is the Maxwell invariant, and $k$ is
an arbitrary rational number. Clearly the special value $k = 1$ corresponds to linear electrodynamics. 

The reason why studying such a class of theories is interesting lies on the fact that Maxwell's theory in higher dimensions $D > 4$ is not conformally invariant, while in a D-dimensional spacetime the electromagnetic stress-energy tensor of the EpM theory is traceless if the power $k$ is chosen to be $k=D/4$. Therefore in four dimensions the linear theory is conformally invariant, and this corresponds of course to the standard Maxwell's theory. In a three-dimensional spacetime, however, if $k=1$ the theory is linear but the electromagnetic stress-energy tensor is not traceless, whereas if $k=3/4$ the theory is conformally invariant but non-linear. Black hole solutions in (1+2)-dimensional and higher-dimensional EpM theories have been obtained in \cite{BH1} and \cite{BH2} respectively, whereas the corresponding (1+2)-dimensional EpM black hole has been studied in the context of scale dependent coupling in \cite{Rincon:2017goj}. In addition, according to \cite{BH2} if $k=0$ one can redefine the parameters to obtain the usual BTZ black hole solution \cite{Banados:1992wn,Banados:1992gq} as well as the corresponding scale dependent counterpart \cite{Koch:2016uso,Rincon:2017ypd}.

In the present article we wish to compute the QNM of charged black holes in non-linear electrodynamics, and in particular in four-dimensional EpM theory.
Our work is organized as follows:  After this Introduction, we present the theory and the corresponding black hole solution in the next Section, while the wave equation for scalar perturbations and QNM are discussed in Section \ref{WE_QNM}. In the fourth Section we compute the QNM of the black holes in the WKB approximation
and we discuss our numerical results. Finally, we conclude our work in Section \ref{Conclusions}.

\section{The theory and the black hole solutions}

We consider the theory in (1+3) dimensions described by the action
\begin{equation}
S[g_{\mu \nu}] = \int \mathrm{d} ^4x \sqrt{-g} \left[ \frac{1}{2 \kappa} R - \alpha (F_{\mu \nu} F^{\mu \nu})^k \right]
\end{equation}
where $\kappa=8 \pi G$, with $G$ being Newton's constant, is the gravitational constant, $k$ is an arbitrary rational number, $R$ is the Ricci scalar, $g$ the determinant of the metric, and $F_{\mu \nu}$ is the electromagnetic field strength. Varying the action with respect to the metric and the gauge field $A_\mu$ one obtains the field equations \cite{BH2}
\begin{eqnarray}
G_{\mu \nu} & = & 4 \kappa \alpha \left [k F_{\mu \rho} F_\nu ^\rho F^{k-1} - \frac{1}{4} g_{\mu \nu} F^k \right ] \\
0 & = & \partial_\mu (\sqrt{-g} F^{\mu \nu} F^{k-1})
\end{eqnarray}
where $F \equiv F_{\mu \nu} F^{\mu \nu}$ is the Maxwell invariant, while $G_{\mu \nu}$ is the Einstein tensor. Seeking spherically symmetric static solutions of the form
\begin{equation}
ds^2 = -f(r) dt^2 + f(r)^{-1} dr^2 + r^2 d \Omega^2
\end{equation}
with $r$ being the radial distance, the metric function $f(r)$ is found to be \cite{BH2}
\begin{equation}
f(r) = 1-\frac{\mu}{r}+\frac{q}{r^\beta}
\end{equation}
where $\mu, q$ are two arbitrary constants related to the mass $M$ and the electric charge $Q$ of the black hole, while the power $\beta$ in four dimensions is
given by \cite{BH2}
\begin{equation}
\beta = \frac{2}{2k-1}
\end{equation}
In the following we shall be considering the case where the power $\beta > 1$.
In this case in order to have real roots for the metric function $f(r)$ the constants $\mu,q$ must satisfy the conditions $\mu > 0$ and also \cite{BH2}
\begin{equation}
0 < q < \left( \frac{\mu}{\beta} \right)^{\beta} (\beta-1)^{\beta-1}
\end{equation}
where the upper bound corresponds to extremal black holes. Clearly the standard Reissner-Nordstr{\"o}m black hole corresponds to $k=1$ and $\beta=2$.
In the following we will compute the QNM of this class of black holes for $\beta=3/2$ (or $k=7/6$) and we shall fix the mass parameter to be $\mu=2$.
We use natural units such that $c = G = \hbar = 1$ and metric signature $(-, +, +, +)$.

\section{The wave equation for scalar perturbations and QNM}\label{WE_QNM}

\begin{figure}[ht!]
\centering
\includegraphics[width=\linewidth]{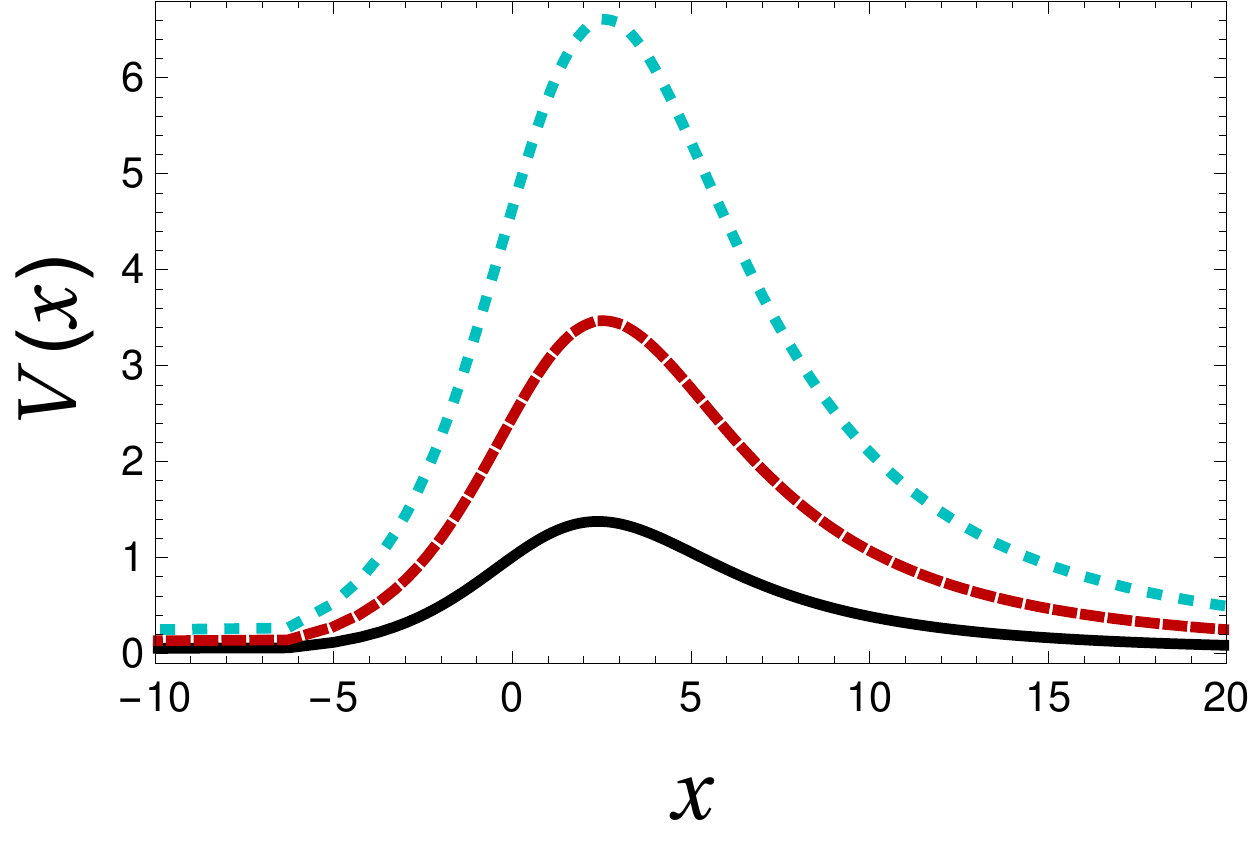}
\caption{Effective potential $V(x)$ as a function of the tortoise coordinate $x$ assuming $\mu = 2$, $q=0.5$ and $\beta=3/2$ for three different cases: i) $l=1$ (solid black line), ii) $l=2$ (dashed red line) and iii) $l=3$ (dotted blue line). Note that the vertical axis
is scaled $1:10$.}
\label{fig:1} 	
\end{figure}

\begin{figure}[ht!]
\centering
\includegraphics[width=\linewidth]{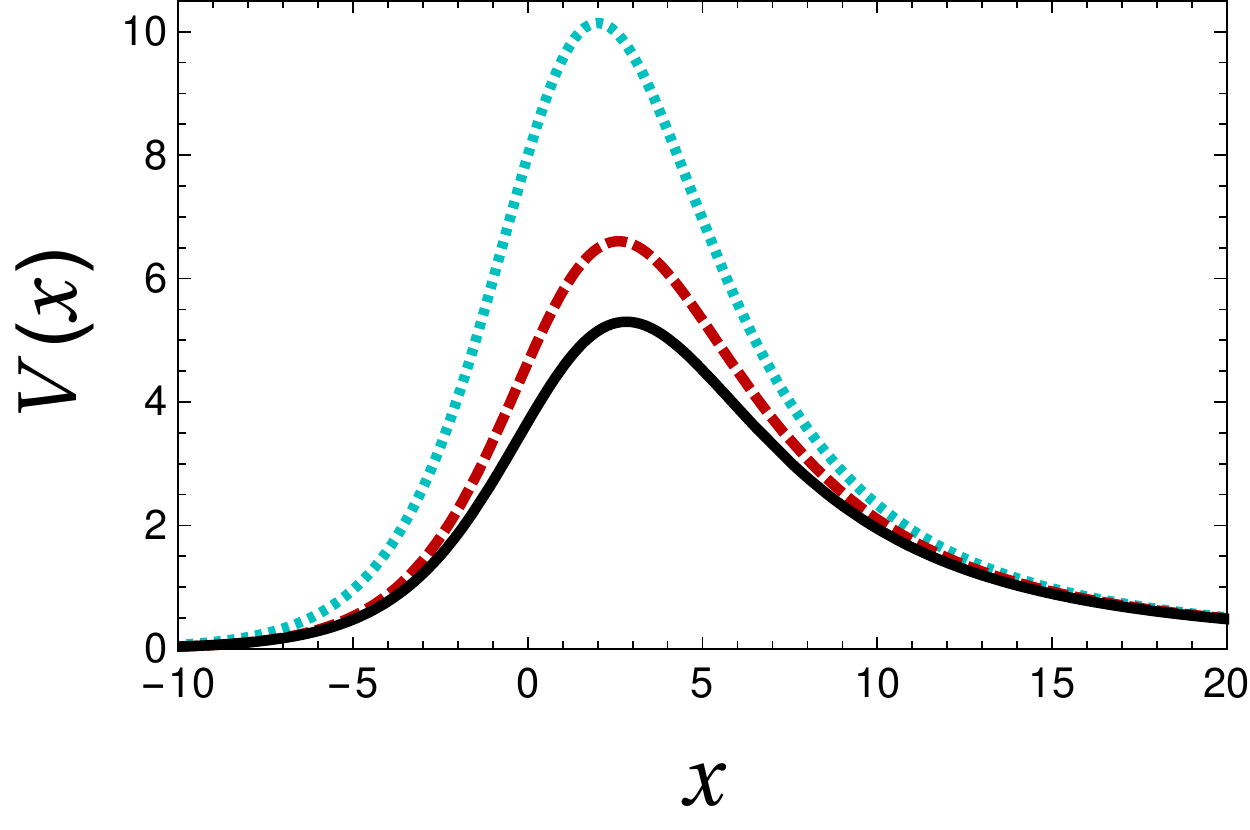}
\caption{Effective potential $V(x)$ as a function of the tortoise coordinate $x$ assuming $\mu = 2$, $l=3$ and $\beta=3/2$ for three different cases: i) $q=0.2$ (solid black line), ii) $q=0.5$ (dashed red line) and iii) $q=0.9$ (dotted blue line). Note that the vertical axis
is scaled $1:10$.}
\label{fig:2} 	
\end{figure}

Next we consider in the above gravitational background a probe minimally coupled massless scalar field with equation of motion
\begin{equation}
\frac{1}{\sqrt{-g}} \partial_\mu (\sqrt{-g} g^{\mu \nu} \partial_\nu) \Phi = 0
\end{equation}
Using the standard ansatz
\begin{equation}\label{separable}
\Phi(t,r,\theta, \phi) = e^{-i \omega t} R(r) Y_l^m (\theta, \phi)
\end{equation}
with $Y_l^m$ being the spherical harmonics, we obtain an ordinary differential equation for the radial part
\begin{equation}
R'' + \left( \frac{2}{r} + \frac{f'}{f} \right) R' + \left( \frac{\omega^2}{f^2} - \frac{l (l+1)}{r^2 f} \right) R = 0
\end{equation}
where the prime denotes differentiation with respect to radial distance $r$, and $l$ is the quantum number of angular momentum. Next we define
new variables as follows
\begin{eqnarray}
R & = & \frac{\psi}{r} \\
x & = & \int \frac{dr}{f(r)}
\end{eqnarray}
with $x$ being the so-called tortoise coordinate, and we recast the equation for the radial part into a Schr{\"o}dinger-like equation of the form
\begin{equation}
\frac{d^2 \psi}{dx^2} + (\omega^2 - V(x)) \psi = 0
\end{equation}
Therefore we obtain for the effective potential the expression
\begin{equation}
V(r) = f(r) \: \left( \frac{l (l+1)}{r^2}+\frac{f'(r)}{r} \right)
\end{equation}
The effective potential as a function of the tortoise coordinate can be seen in Figures \ref{fig:1} and \ref{fig:2}. Fig. \ref{fig:1} shows
how the potential changes with $l$ for a fixed charge, while Fig. \ref{fig:2} shows how the potential changes with the electric charge for a given $l$.

For asymptotically flat spacetimes the Schr{\"o}dinger-like equation is supplemented by the boundary conditions

\begin{equation}
\psi(x) \rightarrow
\left\{
\begin{array}{lcl}
A e^{i \omega x} & \mbox{ if } & x \rightarrow - \infty \\
&
&
\\
 C_+ e^{i \omega x} + C_- e^{-i \omega x} & \mbox{ if } & x \rightarrow + \infty
\end{array}
\right.
\end{equation}
where $A, C_+, C_-$ are arbitrary coefficients. Up to now, the procedure is exactly the same used to compute the so-called graybody factors (GBF), which show the modification of the spectrum of Hawking radiation due to the effective potential barrier, and where the frequency is real and takes continuous values. For an incomplete list see e.g. \cite{col1,col2,col3,col4,col5,col6,col7,3D1,3D2,kanti,Fernando:2004ay,Panotopoulos:2016wuu,coupling,Panotopoulos:2017yoe,kanti2,kanti3,Ahmed:2016lou} and references therein.
The QNM are determined requiring that the first coefficient of the second condition vanishes,
i.e. $C_+ = 0$. The purely ingoing wave physically means that nothing can escape from the horizon, while the purely outgoing wave corresponds
to the requirement that no radiation is incoming from infinity. We thus obtain an infinite set of discrete complex numbers called the quasinormal
frequencies of the black hole.

\section{QNM of charged BH in EpM theory in the WKB approximation}

Computing the QNM of black holes analytically is possible only in very few cases, see e.g. \cite{cardoso2,exact,potential,ferrari}. Semi-analytical methods based on the WKB approximation \cite{wkb1,wkb2} are perhaps the most popular ones, and they have been applied extensively to several cases. For an
incomplete list see e.g. \cite{paper1,paper2,paper3} and for more recent works \cite{paper4,paper5,paper6,paper7,paper8} and references therein. The QN frequencies are given by
\begin{equation}
\omega^2 = V_0+(-2V_0'')^{1/2} \Lambda(n) - i \nu (-2V_0'')^{1/2} [1+\Omega(n)]
\end{equation}
where $n=0,1,2...$ is the overtone number, $\nu=n+1/2$, $V_0$ is the maximum of the effective potential, $V_0''$ is the second derivative of the effective potential evaluated at the maximum, while $\Lambda(n), \Omega(n)$ are complicated expressions of $\nu$ and higher derivatives of the potential evaluated at the maximum, and can be seen e.g. in \cite{paper2,paper7}. Here we have used the Wolfram Mathematica \cite{wolfram} code with WKB at any order from one to six presented in \cite{code}.

In \cite{cardoso3} it was shown that the WKB approximation works very well for $l > n$, and therefore here we shall consider the cases i) ${l=1,n=0}$,
ii) ${l=2,n=0,n=1}$ and iii) ${l=3,n=0,n=1,n=2}$. We have also included the
lowest multiple ${l=0,n=0}$ since it corresponds to the fundamental mode, although we do not expect the approximation to be very good. Finally, we shall consider non-extremal black holes (for the extremal case see \cite{extremal}), while the eikonal regime $l \gg 1$ will be considered separately in the end before concluding our work.

Our numerical results for the QN modes of the EpM black holes are summarized in the tables \ref{table:First set}, \ref{table:Second set}, \ref{table:Third set}, \ref{table:Fourth set}, \ref{table:Fifth set}, \ref{table:Sixth set}, \ref{table:Seventh set}, \ref{table:Eigth set}, \ref{table:Ninth set}, \ref{table:Tenth set} and \ref{table:Eleventh set} below. We have fixed the mass parameter to be $\mu=2$, that corresponds to a black hole mass $M=1$, and for the electromagnetic coupling $q$ we have considered 11 values from 0.10 to 1.07 (the extremal black hole is obtained for $q=1.09$), so that the numerical results go smoothly to the Schwarzschild limit as $q \rightarrow 0$. Finally in the last 2 tables we have made a direct comparison between EpM black holes and the standard Reissner-Nordstr{\"o}m (RN) black hole \cite{RN} for $l=1,n=0$ and $l=2,n=0$.

\begin{figure}[ht!]
\centering
\includegraphics[width=\linewidth]{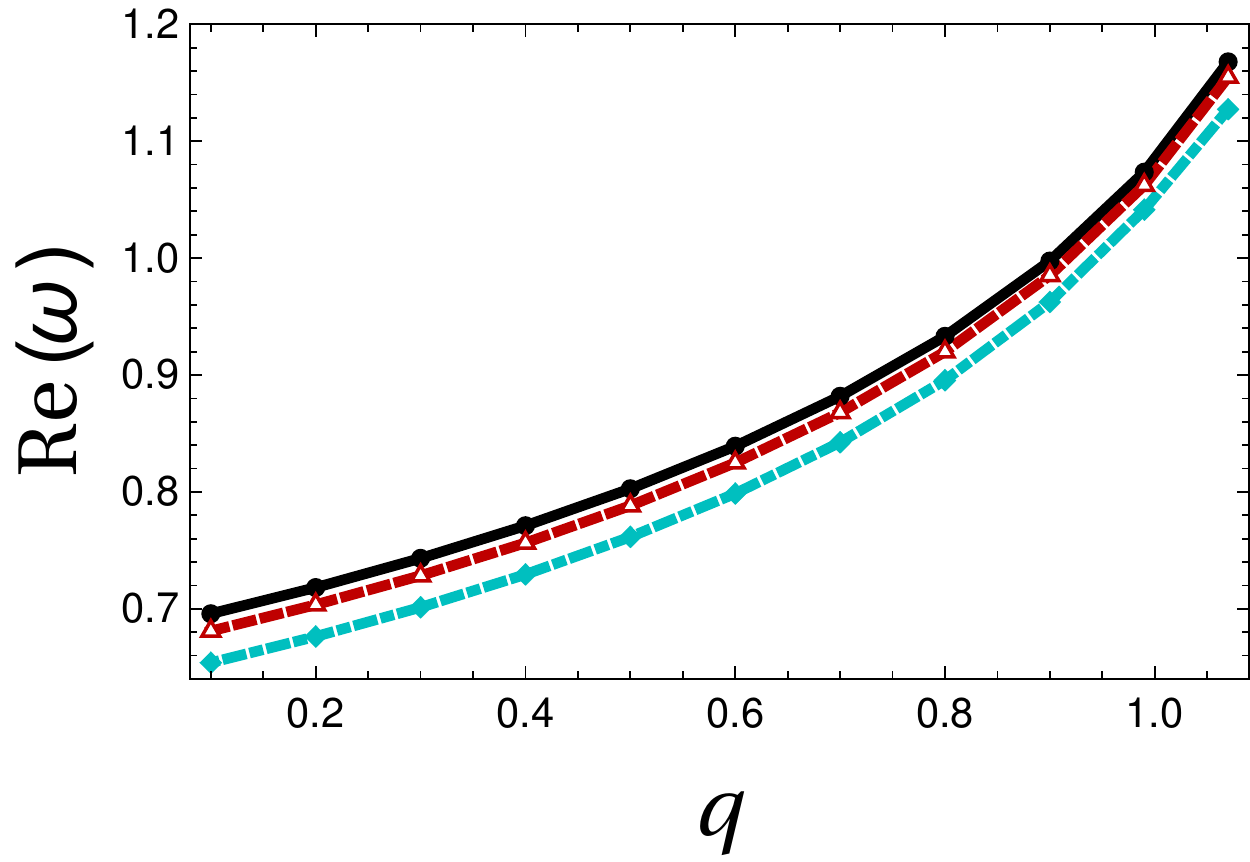}
\caption{Re$(\omega)$ as a function of electromagnetic coupling
$q$ for $l = 3$ in three different cases: i) $n = 0$ (solid black line), ii) $n=1$ (dashed red line) and iii) $n=2$ (dotted-dashed blue line). We have assumed that $\mu = 2$ and $\beta = 3/2$.}
\label{fig:3} 	
\end{figure}

\begin{figure}[ht!]
\centering
\includegraphics[width=\linewidth]{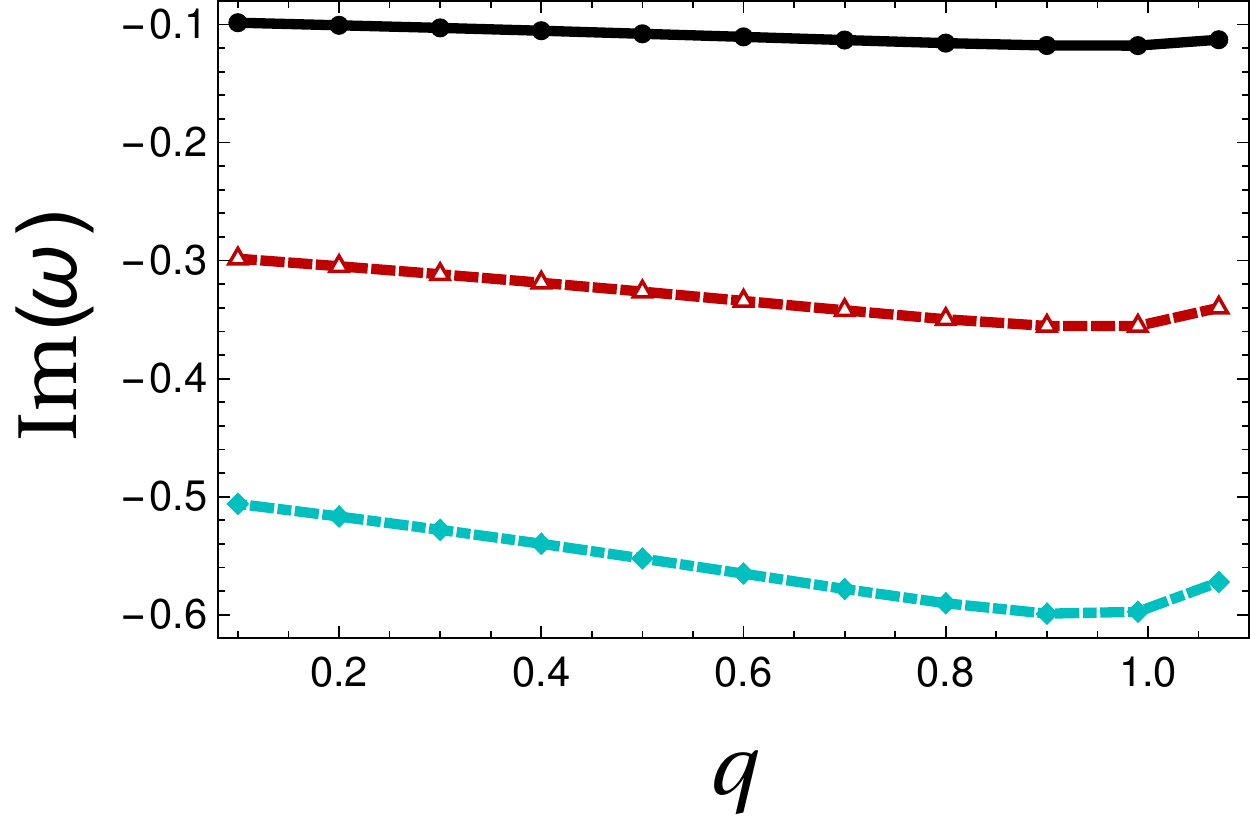}
\caption{Im$(\omega)$ as a function of electromagnetic coupling
$q$ for $l= 3$ in three different cases: i) $n = 0$ (solid black line), ii) $n=1$ (dashed red line) and iii) $n=2$ (dotted-dashed blue line). We have assumed that $\mu = 2$ and $\beta = 3/2$.}
\label{fig:4} 	
\end{figure}

\begin{figure}[ht!]
\centering
\includegraphics[width=\linewidth]{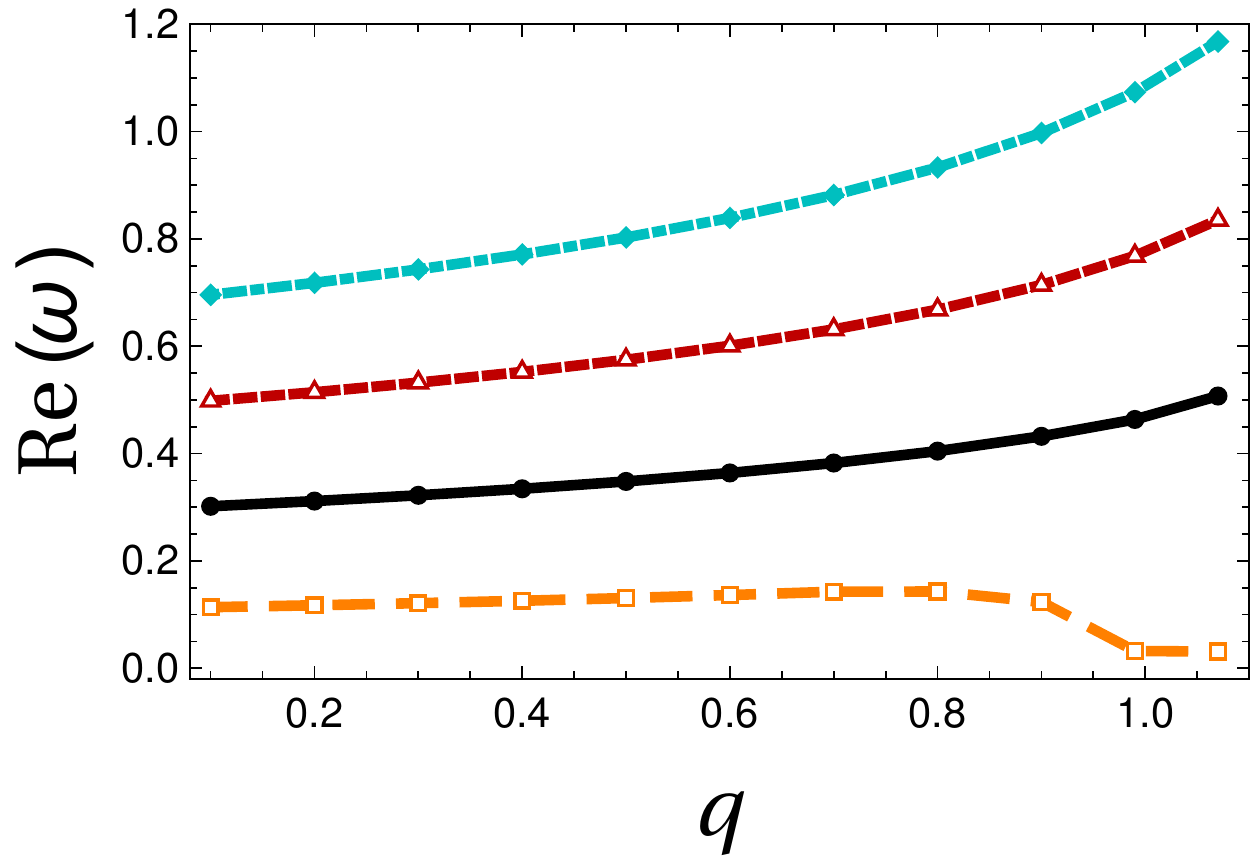}
\caption{Re$(\omega)$ as a function of electromagnetic coupling
$q$ for $n=0$ in four different cases: i) $l = 1$ (solid black line), ii) $l=2$ (short dashed red line) and iii) $l=3$ (dotted-dashed blue line) and iv) $l=0$ (long dashed orange line). We have assumed that $\mu = 2$ and $\beta = 3/2$.}
\label{fig:5} 	
\end{figure}

\begin{figure}[ht!]
\centering
\includegraphics[width=\linewidth]{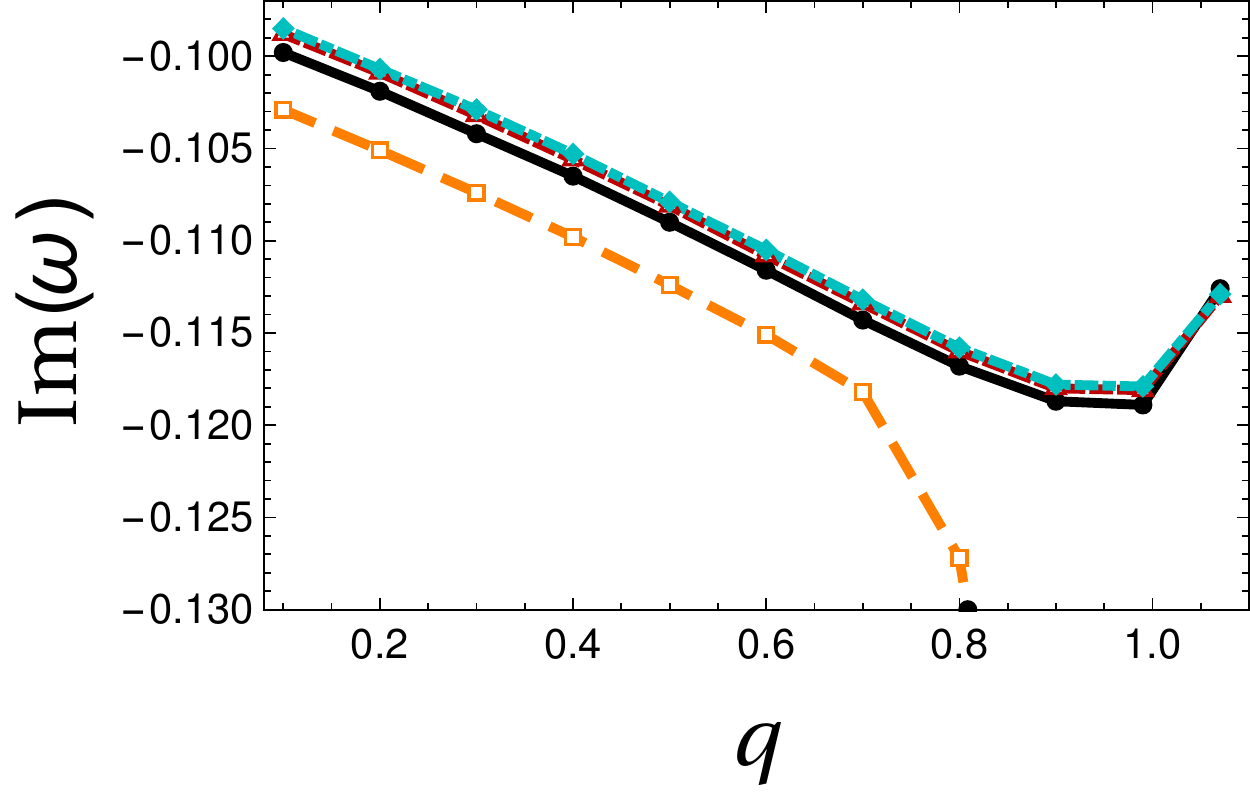}
\caption{Im$(\omega)$ as a function of electromagnetic coupling
$q$ for $n=0$ in three different cases: i) $l = 1$ (solid black line), ii) $l=2$ (short dashed red line) and iii) $l=3$ (dotted-dashed blue line) and iv) $l=0$ (long dashed orange line). We have assumed that $\mu = 2$ and $\beta = 3/2$.}
\label{fig:6} 	
\end{figure}

Next we show graphically how the QNMs depend on the angular momentum, the overtone number as well as the electromagnetic coupling. In particular, Figure \ref{fig:3} shows the real part of the frequencies versus $q$ for 3 different values of the overtone number $n=0,1,2$ and fixed angular momentum $l=3$, while \ref{fig:5} shows the real part of the frequencies versus $q$ for 4 different values of angular momentum $l=0,1,2,3$ and fixed $n=0$. Similarly, Figures \ref{fig:4} and \ref{fig:6} show the imaginary part of the frequencies versus $q$ for the same values of $l,n$. The real part increases with the electric charge and with the angular momentum, like in the case of RN. In other theories too, such as Born-Infeld and Gauss-Bonnet gravity, the real part of QNM of charged BH increases with the electric charge \cite{paper5,paper8}. The imaginary part becomes more and more negative with the electric charge, and less and less negative with the angular momentum, like in the RN case \cite{RN}. It is worth commenting on the characteristic minimum  of the imaginary part (or maximum if one plots $-\text{Im}(\omega)$ versus $q$) for some value of the electric charge close to its maximum value, like in the RN case and contrary to Born-Infeld and Gauss-Bonnet gravity, where the imaginary part is a monotonic function of the electric charge \cite{paper5,paper8}.


Finally, in the eikonal limit ($l \rightarrow \infty$) the WKB method becomes increasingly accurate. As a matter of fact, in the Schwarzschild case it gives the correct result \cite{iyer}, while for near extremal black holes the true potential is given by the P{\"o}schl-Teller (PT) potential \cite{potential,moss,lemos}
\begin{equation}
V(x)=\frac{V_0}{\cosh^2[\alpha (x-x_0)]}
\end{equation}
for which the QN modes can be computed exactly, and in terms of the parameters of the PT potential they are given by \cite{review2,ferrari}
\begin{equation}
\omega^2 = \pm \sqrt{V_0-\left(\frac{1}{2}\alpha\right)^2}-i \alpha \left(n + \frac{1}{2}\right)
\end{equation}
In addition, following the ideas and the formalism developed in \cite{eikonal1}, the QN modes in the eikonal limit can be computed by the formula 
\begin{equation}
\omega_{l \gg 1} = \Omega_c l - i \left(n+\frac{1}{2}\right) |\lambda_L|
\end{equation}
where the Lyapunov exponent $\lambda_L$ is given by \cite{eikonal1}
\begin{equation}
\lambda_L = \sqrt{\frac{1}{2}f(r_c) r_c^2 \left( \frac{d^2}{dr^2} \frac{f}{r^2}  \right)\bigg{|}_{r=r_c}}
\end{equation}
while the angular velocity $\Omega_c$ at the unstable null geodesic is given by \cite{eikonal1}
\begin{equation}
\Omega_c = \sqrt{\frac{f(r_c)}{r_c^2}}
\end{equation}
with $f(r)$ being the lapse metric function, and $r_c$ is the root of the algebraic equation \cite{eikonal1}
\begin{equation}
2 f(r_c) - r_c f'(r)|_{r_c} = 0
\end{equation}
See, however, \cite{ref1} for a counter example where the correspondence between the null geodesics and the eikonal regime does not work. 
For the metric function we consider here, and introducing $z=\sqrt{r}$,
the previous algebraic equation for $r_c$ takes the form of the cubic equation
\begin{equation}
z^3 - \frac{3 \mu z}{2} + \frac{7q}{4} = 0
\end{equation}
the determinant of which is computed to be
\begin{equation}
D = \frac{49q^2}{64}-\frac{\mu^3}{8}
\end{equation}
Given the formulas above for the real part of the frequencies we obtain the expression
\begin{equation}
\omega_R^L = l \frac{\sqrt{\mu-\frac{3q}{2z}}}{\sqrt{2} \left( \frac{3 \mu z}{2}-\frac{7q}{4} \right)}
\end{equation}
while for the imaginary part we obtain the expression
\begin{equation}
\omega_I^L = -\left( n+\frac{1}{2} \right) \; \left| \frac{\omega_R}{l z \sqrt{2}} \sqrt{-3 \mu + \frac{21q}{4z}} \right|
\end{equation}
where $z$ is given by
\begin{equation}
z = \left( -\frac{7q}{8}-\sqrt{\frac{49 q^2}{64}-\frac{\mu^3}{8}} \right)^{1/3} + \left( -\frac{7q}{8} + \sqrt{\frac{49 q^2}{64}-\frac{\mu^3}{8}} \right)^{1/3}
\end{equation}
The formulas above are valid in linear electrodynamics. In the presence of non-linear electromagnetic fields, however, photons follow the null geodesics of an effective geometry \cite{eikonal2,eikonal3,ref2}
\begin{equation}
ds_{eff}^2 = \frac{G_m}{G_e} [-f(r) dt^2 + f(r)^{-1} dr^2 + r^2 d \Omega_2^2] 
\end{equation}
where the magnetic factor $G_m=1$ if the BH is not magnetically charged, while
the electric factor $G_e$ is given by \cite{eikonal2,eikonal3}
\begin{equation}
G_e = 1 - 4 \mathcal{L}_{FF} \frac{q_e^2}{\mathcal{L}_F^3 r^4} 
\end{equation}
with $\mathcal{L}_F$ and $\mathcal{L}_{FF}$ being the first and the second derivative respectively of the electromagnetic Lagrangian with respect to the Maxwell invariant, while the electric charge $q_e$ is given by
\begin{equation}
q_e = r^2 \mathcal{L}_F F^{rt}
\end{equation}
In this framework the expressions valid in linear electrodynamics are modified by factors of $G_m/G_e$ and derivatives with respect to $r$ evaluated at $r_c$. 
For the EpM theory $\mathcal{L}=F^k$, however, the factor $G_e$ turns out to be a constant, $G_e=2k-1=4/3$, which simplifies things. In this case the algebraic equation for $r_c$ as well as the Lyapunov exponent remain the same \cite{eikonal2,eikonal3}, while the real part of the frequencies is corrected by a pre-factor $1/\sqrt{G_e}$ \cite{eikonal2,eikonal3}.
Therefore finally we obtain the following expressions for the QN modes in the eikonal limit in EpM theory
\begin{eqnarray}
\omega_R^{EpM} & = & \frac{l}{\sqrt{2 G_e}} \frac{\sqrt{\mu-\frac{3q}{2z}}}{ \left( \frac{3 \mu z}{2}-\frac{7q}{4} \right)} \\
\omega_I^{EpM} & = & -\left( n+\frac{1}{2} \right) \; \left| \sqrt{\frac{G_e}{2}} \frac{\omega_R}{l z} \sqrt{-3 \mu + \frac{21q}{4z}} \right|
\end{eqnarray}


\begin{table}[ph]
\tbl{QN frequencies for electric charge $q=0.10$.}
{
\begin{tabular}{@{}ccccc@{}} \toprule
$n$ & $l=0$ & $l=1$ & $l=2$ & $l=3$ 
\\ \colrule
0  &  0.1138-0.1029 i  & 0.3018-0.0998 i  & 0.4983-0.0988 i & 0.6958-0.0985 i \\
1 &  &  & 0.4784-0.3017 i & 0.6810-0.2984 i  \\
2 &  &  &  & 0.6539-0.5061 i
\\ \botrule
\end{tabular} 
\label{table:First set}
}
\end{table}

\begin{table}[ht!]
\tbl{QN frequencies for electric charge $q=0.20$.}
{
\begin{tabular}{@{}ccccc@{}} \toprule
$n$ & $l=0$ & $l=1$ & $l=2$ & $l=3$ 
\\ \colrule
0  & 0.1173-0.1051 i & 0.3115-0.1019 i & 0.5143-0.1009 i & 0.7182-0.1007 i \\
1 &  &  & 0.4945-0.3081 i & 0.7035-0.3048 i \\
2 &  &  &  & 0.6763-0.5168 i
\\ \botrule
\end{tabular} 
\label{table:Second set}
}
\end{table}

\begin{table}[ht!]
\tbl{QN frequencies for electric charge $q=0.30$.}
{
\begin{tabular}{@{}ccccc@{}} \toprule
$n$ & $l=0$ & $l=1$ & $l=2$ & $l=3$ 
\\ \colrule
0  & 0.1213-0.1074 i & 0.3223-0.1042 i & 0.5322-0.1032 i & 0.7432-0.1029 i  \\
1 &  &  & 0.5124-0.3149 i & 0.7284-0.3116 i  \\
2 &  &  &  & 0.7013-0.5281 i
\\ \botrule
\end{tabular} 
\label{table:Third set}
}
\end{table}

\begin{table}[ht!]
\tbl{QN frequencies for electric charge $q=0.40$.}
{
\begin{tabular}{@{}ccccc@{}} \toprule
$n$ & $l=0$ & $l=1$ & $l=2$ & $l=3$ 
\\ \colrule
0  & 0.1257-0.1098 i & 0.3344-0.1065 i & 0.5522-0.1056 i & 0.7711-0.1053 i  \\
1 &  &  & 0.5324-0.3220 i & 0.7564-0.3188 i  \\
2 &  &  & & 0.7294-0.5399 i
\\ \botrule
\end{tabular} 
\label{table:Fourth set}
}
\end{table}

\begin{table}[ht!]
\tbl{QN frequencies for electric charge $q=0.50$.}
{
\begin{tabular}{@{}ccccc@{}} \toprule
$n$ & $l=0$ & $l=1$ & $l=2$ & $l=3$ 
\\ \colrule
0  & 0.1308-0.1124 i & 0.3481-0.1090 i & 0.5748-0.1081 i & 0.8027-0.1079 i  \\
1 &  &  & 0.5553-0.3295 i & 0.7882-0.3263 i  \\
2 &  &  &  & 0.7615-0.5524 i
\\ \botrule
\end{tabular} 
\label{table:Fifth set}
}
\end{table}

\begin{table}[ht!]
\tbl{QN frequencies for electric charge $q=0.60$.}
{
\begin{tabular}{@{}ccccc@{}} \toprule
$n$ & $l=0$ & $l=1$ & $l=2$ & $l=3$ 
\\ \colrule
0  & 0.1364-0.1151 i & 0.3639-0.1116 i  & 0.6009-0.1108 i  & 0.8391-0.1105 i  \\
1  &  &  & 0.5817-0.3373 i & 0.8249-0.3342 i  \\
2  &  &  &  & 0.7986-0.5652 i
\\ \botrule
\end{tabular} 
\label{table:Sixth set}
}
\end{table}

\begin{table}[ht!]
\tbl{QN frequencies for electric charge $q=0.70$.}
{
\begin{tabular}{@{}ccccc@{}} \toprule
$n$ & $l=0$ & $l=1$ & $l=2$ & $l=3$ 
\\ \colrule
0  & 0.1426-0.1182 i & 0.3824-0.1143 i  & 0.6314-0.1134 i  & 0.8818-0.1132 i  \\
1  &  &  & 0.6128-0.3451 i & 0.8679-0.3421 i  \\
2  &  &  &  & 0.8424-0.5782 i
\\ \botrule
\end{tabular} 
\label{table:Seventh set}
}
\end{table}

\begin{table}[ht!]
\tbl{QN frequencies for electric charge $q=0.80$.}
{
\begin{tabular}{@{}ccccc@{}} \toprule
$n$ & $l=0$ & $l=1$ & $l=2$ & $l=3$ 
\\ \colrule
0  & 0.1430-0.1272 i & 0.4046-0.1168 i  & 0.6682-0.1160 i  & 0.9331-0.1158 i  \\
1  &  &  & 0.6504-0.3525 i & 0.9199-0.3497 i  \\
2  &  &  &  & 0.8955-0.5903 i
\\ \botrule
\end{tabular} 
\label{table:Eigth set}
}
\end{table}

\begin{table}[ht!]
\tbl{QN frequencies for electric charge $q=0.90$.}
{
\begin{tabular}{@{}ccccc@{}} \toprule
$n$ & $l=0$ & $l=1$ & $l=2$ & $l=3$ 
\\ \colrule
0  & 0.1234-0.1595 i & 0.4322-0.1187 i  & 0.7142-0.1180 i  & 0.9975-0.1178 i  \\
1  &  &  & 0.6976-0.3579 i & 0.9852-0.3554 i  \\
2  &  &  &  & 0.9624-0.5991 i
\\ \botrule
\end{tabular} 
\label{table:Ninth set}
}
\end{table}

\begin{table}[ht!]
\tbl{QN frequencies for electric charge $q=0.99$.}
{
\begin{tabular}{@{}ccccc@{}} \toprule
$n$ & $l=0$ & $l=1$ & $l=2$ & $l=3$ 
\\ \colrule
0  & 0.0322-0.6329 i & 0.4636-0.1189 i  & 0.7684-0.1181 i  & 1.0735-0.1179 i  \\
1  &  &  & 0.7529-0.3575 i & 1.0623-0.3554 i  \\
2  &  &  &  & 1.0414-0.5975 i
\\ \botrule
\end{tabular} 
\label{table:Tenth set}
}
\end{table}

\begin{table}[ht!]
\tbl{QN frequencies for electric charge $q=1.07$.}
{
\begin{tabular}{@{}ccccc@{}} \toprule
$n$ & $l=0$ & $l=1$ & $l=2$ & $l=3$ 
\\ \colrule
0  & 0.0310-0.5685 i & 0.5073-0.1126 i  & 0.8354-0.1130 i  & 1.1679-0.1129 i  \\
1  &  &  & 0.8202-0.3405 i & 1.1549-0.3400 i  \\
2  &  &  &  & 1.1273-0.5723 i
\\ \botrule
\end{tabular} 
\label{table:Eleventh set}
}
\end{table}

\begin{table}[ht!]
\tbl{QN frequencies for RN and EpM black holes for $n=0$ and $l=1$.}
{
\begin{tabular}{@{}ccc@{}} \toprule
$q$ & $RN$ & $EpM$ 
\\ \colrule
0.10  & 0.2980-0.0982 i & 0.3018-0.0998 i   \\
0.20  & 0.3036-0.0987 i &  0.3115-0.1019 i  \\
0.30  & 0.3096-0.0990 i & 0.3223-0.1042 i  \\
0.40  & 0.3162-0.0993 i & 0.3344-0.1065 i  \\
0.50  & 0.3235-0.0994 i & 0.3481-0.1090 i  \\
0.60  & 0.3317-0.0993 i & 0.3639-0.1116 i  \\
0.70  & 0.3409-0.0987 i & 0.3824-0.1143 i  \\
0.80  & 0.3515-0.0974 i & 0.4046-0.1168 i  \\
0.90  & 0.3639-0.0947 i & 0.4322-0.1187 i  \\
0.95 & 0.3706-0.0925 i & 0.4491-0.1190 i  \\
0.99 & 0.3766-0.0900 i & 0.4636-0.1189 i
\\ \botrule
\end{tabular} 
\label{table:Penultimo set}
}
\end{table}

\begin{table}[ht!]
\tbl{QN frequencies for RN and EpM black holes for $n=0$ and $l=2$.}
{
\begin{tabular}{@{}ccc@{}} \toprule
$q$ & $RN$ & $EpM$ 
\\ \colrule
0.10  & 0.4920-0.0973 i & 0.4983-0.0988 i   \\
0.20  & 0.5011-0.0977 i &  0.5143-0.1009 i  \\
0.30  & 0.5110-0.0981 i & 0.5322-0.1032 i  \\
0.40  & 0.5218-0.0984 i & 0.5522-0.1056 i  \\
0.50  & 0.5338-0.0986 i & 0.5748-0.1081 i  \\
0.60  & 0.5472-0.0985 i & 0.6009-0.1108 i  \\
0.70  & 0.5624-0.0980 i & 0.6314-0.1134 i  \\
0.80  & 0.5800-0.0968 i & 0.6682-0.1160 i  \\
0.90  & 0.6010-0.0943 i & 0.7142-0.1180 i  \\
0.95 & 0.6131-0.0921 i & 0.7423-0.1184 i  \\
0.99 & 0.6238-0.0895 i & 0.7684-0.1181 i 
\\ \botrule
\end{tabular} 
\label{table:Last set}
}
\end{table}

\section{Conclusions}\label{Conclusions}

In this article we have computed the quasinormal modes of four-dimensional charged black holes in Einstein-power-Maxwell non-linear electrodynamics.
We have studied scalar perturbations using a Schr{\"o}dinger-like equation with the appropriate effective potential, and we have adopted the popular and extensively used WKB approximation. Our numerical results are summarized in tables, and we have shown graphically the behaviour of the spectrum on the electric charge of the black hole as well as on the overtone number and on the quantum number of angular momentum. An analytical expression for the QN spectrum in the eikonal regime has been obtained, and the comparison with the RN black hole as well as with charged black holes in other theories, such as Born-Infeld and Gauss-Bonnet gravity, is made.


\section*{Acknowlegements}

We are grateful to the anonymous reviewer for valuable suggestions that significantly improved the quality of our paper. The author G. P. thanks the Funda\c c\~ao para a Ci\^encia e Tecnologia (FCT), Portugal, for the financial support to the Multidisciplinary Center for Astrophysics (CENTRA),  Instituto Superior T\'ecnico,  Universidade de Lisboa, through the
Grant No. UID/FIS/00099/2013. The author A. R. was supported
by the CONICYT-PCHA/\- Doctorado Nacional/2015-21151658. Finally we wish to thank
R. A. Konoplya and A. Zhidenko for communications. 



\begin{thebibliography}{99}
\bibitem{wheeler} T.~Regge and J.~A.~Wheeler,
  Phys.\ Rev.\  {\bf 108} (1957) 1063.

\bibitem{zerilli1} F.~J.~Zerilli,
  Phys.\ Rev.\ Lett.\  {\bf 24} (1970) 737.
  
\bibitem{zerilli2} F.~J.~Zerilli,
  Phys.\ Rev.\ D {\bf 2} (1970) 2141.
  
\bibitem{zerilli3} F.~J.~Zerilli,
  Phys.\ Rev.\ D {\bf 9} (1974) 860.

\bibitem{moncrief} V.~Moncrief,
  Phys.\ Rev.\ D {\bf 12} (1975) 1526.

\bibitem{teukolsky} S.~A.~Teukolsky,
  Phys.\ Rev.\ Lett.\  {\bf 29} (1972) 1114.

\bibitem{monograph} S.~Chandrasekhar,
  ``The mathematical theory of black holes,''
  OXFORD, UK: CLARENDON (1985) 646 P.

\bibitem{ligo1} B.~P.~Abbott {\it et al.} [LIGO Scientific and Virgo Collaborations],
  Phys.\ Rev.\ Lett.\  {\bf 116} (2016) no.6,  061102
[arXiv:1602.03837 [gr-qc]].

\bibitem{ligo2} B.~P.~Abbott {\it et al.} [LIGO Scientific and Virgo Collaborations],
  Phys.\ Rev.\ Lett.\  {\bf 116} (2016) no.24,  241103
[arXiv:1606.04855 [gr-qc]].

\bibitem{ligo3} B.~P.~Abbott {\it et al.} [LIGO Scientific and VIRGO Collaborations],
  Phys.\ Rev.\ Lett.\  {\bf 118} (2017) no.22,  221101
[arXiv:1706.01812 [gr-qc]].

\bibitem{review1} K.~D.~Kokkotas and B.~G.~Schmidt,
  Living Rev.\ Rel.\  {\bf 2} (1999) 2
[gr-qc/9909058].

\bibitem{review2} E.~Berti, V.~Cardoso and A.~O.~Starinets,
  Class.\ Quant.\ Grav.\  {\bf 26} (2009) 163001
  [arXiv:0905.2975 [gr-qc]].

\bibitem{BI} M.~Born and L.~Infeld,
��Proc.\ Roy.\ Soc.\ Lond.\ A {\bf 144} (1934) 425.

\bibitem{ST1} M. B. Green, J. H. Schwarz and E. Witten, \textit{Superstring Theory, Vol. 1 \& 2},
Cambridge Monographs on Mathematical Physics.

\bibitem{ST2} J. Polchinski, \textit{String Theory, Vol. 1 \& 2}, Cambridge Monographs on Mathematical Physics.

\bibitem{Dbranes1} C. V. Johnson, \textit{D-Branes}, Cambridge Monographs on Mathematical Physics.

\bibitem{Dbranes2} B. Zwiebach, \textit{A First Course in String Theory }, Cambridge University Press.

\bibitem{BH1} O.~Gurtug, S.~H.~Mazharimousavi and M.~Halilsoy,
  Phys.\ Rev.\ D {\bf 85} (2012) 104004
[arXiv:1010.2340 [gr-qc]].

\bibitem{BH2} M.~Hassaine and C.~Martinez,
  Class.\ Quant.\ Grav.\  {\bf 25} (2008) 195023
[arXiv:0803.2946 [hep-th]].

\bibitem{Rincon:2017goj} 
  \'A.~Rinc\'on, E.~Contreras, P.~Bargue\~no, B.~Koch, G.~Panotopoulos and A.~Hern\'andez-Arboleda,
  Eur.\ Phys.\ J.\ C {\bf 77}, no. 7, 494 (2017)
[arXiv:1704.04845 [hep-th]].
  
\bibitem{Banados:1992wn}
  M.~Banados, C.~Teitelboim and J.~Zanelli,
  Phys.\ Rev.\ Lett.\  {\bf 69}, 1849 (1992)
[hep-th/9204099].

\bibitem{Banados:1992gq}
  M.~Banados, M.~Henneaux, C.~Teitelboim and J.~Zanelli,
  Phys.\ Rev.\ D {\bf 48}, 1506 (1993)
  Erratum: [Phys.\ Rev.\ D {\bf 88}, 069902 (2013)]
[gr-qc/9302012].

\bibitem{Koch:2016uso} B.~Koch, I.~A.~Reyes and \'A.~Rinc\'on,
  Class.\ Quant.\ Grav.\  {\bf 33}, no. 22, 225010 (2016)
[arXiv:1606.04123 [hep-th]].

\bibitem{Rincon:2017ypd} \'A.~Rinc\'on, B.~Koch and I.~Reyes,
  J.\ Phys.\ Conf.\ Ser.\  {\bf 831}, no. 1, 012007 (2017)
[arXiv:1701.04531 [hep-th]].

\bibitem{col1} C. Doran,  A. Lasenby,  S. Dolan,  and  I.  Hinder,  Phys.
Rev. D {\bf 71}, 124020 (2005).

\bibitem{col2} S. Dolan, C. Doran, and A. Lasenby, Phys. Rev. D {\bf 74}, 064005 (2006).

\bibitem{col3} L. C. B. Crispino, E. S. Oliveira, A. Higuchi, and G. E. A. Matsas,
Phys. Rev. D {\bf 75}, 104012 (2007).

\bibitem{col4} S.  R.  Dolan,  Classical  Quantum  Gravity {\bf 25}, 235002 (2008).

\bibitem{col5} L. C. B. Crispino, S. R. Dolan, and E. S. Oliveira,
Phys. Rev. Lett. {\bf 102}, 231103 (2009).

\bibitem{col6} L. C. B. Crispino, S. R. Dolan, and E. S. Oliveira,
Phys. Rev. D {\bf 79}, 064022 (2009).

\bibitem{col7} L. B. Crispino, A. Higuchi, and E. S. Oliveira,
Phys. Rev. D {\bf 80}, 104026 (2009).

\bibitem{3D1} D.~Birmingham, I.~Sachs and S.~Sen,
  Phys.\ Lett.\ B {\bf 413} (1997) 281
  [hep-th/9707188].
  
\bibitem{3D2} Y.~S.~Myung,
  Mod.\ Phys.\ Lett.\ A {\bf 18} (2003) 617
  [hep-th/0201176].

\bibitem{kanti} P.~Kanti and J.~March-Russell,
  Phys.\ Rev.\ D {\bf 66} (2002) 024023
  [hep-ph/0203223].

\bibitem{Fernando:2004ay} S.~Fernando,
  Gen.\ Rel.\ Grav.\  {\bf 37} (2005) 461,
  [hep-th/0407163].

\bibitem{Panotopoulos:2016wuu} 
  G.~Panotopoulos and \'A.~Rinc\'on,
  Phys.\ Lett.\ B {\bf 772}, 523 (2017)
[arXiv:1611.06233 [hep-th]].
  
\bibitem{coupling} L.~C.~B.~Crispino, A.~Higuchi, E.~S.~Oliveira and J.~V.~Rocha,
  Phys.\ Rev.\ D {\bf 87} (2013) 104034
  [arXiv:1304.0467 [gr-qc]].

\bibitem{Panotopoulos:2017yoe} G.~Panotopoulos and \'A.~Rinc\'on,
  Phys.\ Rev.\ D {\bf 96}, no. 2, 025009 (2017)
[arXiv:1706.07455 [hep-th]].

\bibitem{kanti2} P.~Kanti, T.~Pappas and N.~Pappas,
  Phys.\ Rev.\ D {\bf 90} (2014) no.12,  124077
  [arXiv:1409.8664 [hep-th]];

\bibitem{kanti3} T.~Pappas, P.~Kanti and N.~Pappas,
  Phys.\ Rev.\ D {\bf 94} (2016) no.2,  024035
  [arXiv:1604.08617 [hep-th]].

\bibitem{Ahmed:2016lou} J.~Ahmed and K.~Saifullah,
  arXiv:1610.06104 [gr-qc].

\bibitem{cardoso2} V.~Cardoso and J.~P.~S.~Lemos,
  Phys.\ Rev.\ D {\bf 63} (2001) 124015
[gr-qc/0101052].

\bibitem{exact} D.~Birmingham,
  Phys.\ Rev.\ D {\bf 64} (2001) 064024
[hep-th/0101194].

\bibitem{potential} G.~Poschl and E.~Teller,
  Z.\ Phys.\  {\bf 83} (1933) 143.

\bibitem{ferrari} V.~Ferrari and B.~Mashhoon,
  Phys.\ Rev.\ D {\bf 30} (1984) 295.

\bibitem{wkb1} S.~Iyer and C.~M.~Will,
  Phys.\ Rev.\ D {\bf 35} (1987) 3621.

\bibitem{wkb2} R.~A.~Konoplya,
  Phys.\ Rev.\ D {\bf 68} (2003) 024018
[gr-qc/0303052].

\bibitem{paper1} S.~Iyer,
  Phys.\ Rev.\ D {\bf 35} (1987) 3632.

\bibitem{paper2} K.~D.~Kokkotas and B.~F.~Schutz,
  Phys.\ Rev.\ D {\bf 37} (1988) 3378.

\bibitem{paper3} E.~Seidel and S.~Iyer,
  Phys.\ Rev.\ D {\bf 41} (1990) 374.

\bibitem{paper4} V.~Santos, R.~V.~Maluf and C.~A.~S.~Almeida,
  Phys.\ Rev.\ D {\bf 93} (2016) no.8,  084047
[arXiv:1509.04306 [gr-qc]].

\bibitem{paper5} S.~Fernando and C.~Holbrook,
  Int.\ J.\ Theor.\ Phys.\  {\bf 45} (2006) 1630
  doi:10.1007/s10773-005-9024-9
  [hep-th/0501138].

\bibitem{paper6} J.~L.~Blázquez-Salcedo, F.~S.~Khoo and J.~Kunz,
  arXiv:1706.03262 [gr-qc].

\bibitem{paper7} S.~K.~Chakrabarti,
  Gen.\ Rel.\ Grav.\  {\bf 39} (2007) 567
[hep-th/0603123].

\bibitem{paper8} R.~Konoplya,
  Phys.\ Rev.\ D {\bf 71} (2005) 024038
[hep-th/0410057].

\bibitem{wolfram} 
\url{http://www.wolfram.com}

\bibitem{code} R.~A.~Konoplya and A.~Zhidenko,
  Phys.\ Rev.\ D {\bf 81} (2010) 124036
  [arXiv:1004.1284 [hep-th]].

\bibitem{cardoso3}  V.~Cardoso, J.~P.~S.~Lemos and S.~Yoshida,
  Phys.\ Rev.\ D {\bf 69} (2004) 044004
[gr-qc/0309112].

\bibitem{extremal} H.~Onozawa, T.~Mishima, T.~Okamura and H.~Ishihara,
  Phys.\ Rev.\ D {\bf 53} (1996) 7033
[gr-qc/9603021].

\bibitem{RN} R.~A.~Konoplya,
  Phys.\ Rev.\ D {\bf 66} (2002) 084007
[gr-qc/0207028].

\bibitem{iyer} S.~Iyer,
  Phys.\ Rev.\ D {\bf 35} (1987) 3632.

\bibitem{moss} I.~G.~Moss and J.~P.~Norman,
  Class.\ Quant.\ Grav.\  {\bf 19} (2002) 2323
[gr-qc/0201016].

\bibitem{lemos} V.~Cardoso and J.~P.~S.~Lemos,
  Phys.\ Rev.\ D {\bf 67} (2003) 084020
[gr-qc/0301078].

\bibitem{eikonal1} V.~Cardoso, A.~S.~Miranda, E.~Berti, H.~Witek and V.~T.~Zanchin,
  Phys.\ Rev.\ D {\bf 79} (2009) 064016
  [arXiv:0812.1806 [hep-th]].
  
\bibitem{ref1} R.~A.~Konoplya and Z.~Stuchlík,
  Phys.\ Lett.\ B {\bf 771} (2017) 597
  [arXiv:1705.05928 [gr-qc]].

\bibitem{eikonal2} N.~Bret{\'o}n and L.~A.~Lopez,
  Phys.\ Rev.\ D {\bf 94} (2016) no.10,  104008
  [arXiv:1607.02476 [gr-qc]].

\bibitem{eikonal3} N.~Bret{\'o}n, T.~Clark and S.~Fernando,
  Int.\ J.\ Mod.\ Phys.\ D {\bf 26} (2017) no.10,  1750112
  [arXiv:1703.10070 [gr-qc]].
  
\bibitem{ref2} E.~Chaverra, J.~C.~Degollado, C.~Moreno and O.~Sarbach,
  Phys.\ Rev.\ D {\bf 93} (2016) no.12,  123013
  [arXiv:1605.04003 [gr-qc]].
\end{thebibliography}
\end{document}